\documentclass[a4paper]{jpconf}
\usepackage{graphicx}
\usepackage{hyperref}
\begin{document}
\title{KCDC -- The KASCADE Cosmic-ray Data Centre}
% Search for diffuse gamma-rays

\author{
A~Haungs$^{1}$,
J~Bl\"umer$^{1,2}$, 
B~Fuchs$^{2,a}$,
D~Kang$^{2}$, 
S~Schoo$^{1}$,
D~Wochele$^{1}$,
J~Wochele$^{1}$,
W~D~Apel$^{1}$, 
J~C~Arteaga-Vel\'azquez$^{3}$,
K~Bekk$^{1}$,
M~Bertaina$^{4}$,
H~Bozdog$^{1}$,
I~M~Brancus$^{5}$,
E~Cantoni$^{4,6,b}$, 
A~Chiavassa$^{4}$,
F~Cossavella$^{2,c}$,
K~Daumiller$^{1}$,
V~de Souza$^{7}$,
F~Di~Pierro$^{4}$,
P~Doll$^{1}$,
R~Engel$^{1}$,
D~Fuhrmann$^{8,d}$,
A~Gherghel-Lascu$^{5}$,
H~J~Gils$^{1}$,
R~Glasstetter$^{8}$,
C~Grupen$^{9}$,
D~Heck$^{1}$,
J~R~H\"orandel$^{10}$,
D~Huber$^{2}$,
T~Huege$^{1}$,
K~H~Kampert$^{8}$,
H~O~Klages$^{1}$,
K~Link$^{2}$,
P~{\L}uczak$^{11}$,
H~J~Mathes$^{1}$,
H~J~Mayer$^{1}$,
J~Milke$^{1}$,
B~Mitrica$^{5}$,
C~Morello$^{6}$,
J~Oehlschl\"ager$^{1}$,
S~Ostapchenko$^{12}$,
N~Palmieri$^{2}$,
M~Petcu$^{5}$,
T~Pierog$^{1}$,
H~Rebel$^{1}$,
M~Roth$^{1}$,
H~Schieler$^{1}$,
F~G~Schr\"oder$^{1}$,
O~Sima$^{13}$,
G~Toma$^{5}$,
G~C~Trinchero$^{6}$,
H~Ulrich$^{1}$,
A~Weindl$^{1}$,
J~Zabierowski$^{11}$
}

\address{
$^1$ Institut f\"ur Kernphysik, KIT - Karlsruher Institut f\"ur Technologie, Germany\\
$^2$ Institut f\"ur Experimentelle Kernphysik, KIT - Karlsruher Institut f\"ur Technologie, Germany\\
$^3$ Universidad Michoacana, Instituto de Fisica y Matem\'aticas, Mexico\\
$^4$ Dipartimento di Fisica, Universit\`a degli Studi di Torino, Italy\\
$^5$ National Institute of Physics and Nuclear Engineering, Bucharest, Romania\\
$^6$ Istituto di Fisica dello Spazio Interplanetario, INAF Torino, Italy\\
$^7$ Universidade S\~ao Paulo, Instituto de F\'{\i}sica de S\~ao Carlos, Brasil\\
$^8$ Fachbereich Physik, Universit\"at Wuppertal, Germany\\
$^9$ Fachbereich Physik, Universit\"at Siegen, Germany\\
$^{10}$ Dept. of Astrophysics, Radboud University Nijmegen, The Netherlands\\
$^{11}$ National Centre for Nuclear Research, Department of Astrophysics, {\L}\'{o}d\'{z}, Poland\\
$^{12}$ Frankfurt Institute for Advanced Studies (FIAS), Frankfurt am Main, Germany\\
$^{13}$ Department of Physics, University of Bucharest, Bucharest, Romania\\
\scriptsize{
$^{a}$ now at: DLR Stuttgart, Germany; \\
$^{b}$ now at: Istituto Nazionale di Ricerca Metrologia, INRIM, Torino;\\
$^{c}$ now at: DLR Oberpfaffenhofen, Germany; \\
$^{d}$ now at: University of Duisburg-Essen, Duisburg, Germany\\
}
}
\ead{haungs@kit.edu}

\begin{abstract}
KCDC, the `KASCADE Cosmic-ray Data Centre', is a web portal, where data of astroparticle
physics experiments will be made available for the interested public. The KASCADE
experiment, financed by public money, was a large-area detector for the measurement of 
high-energy cosmic rays via the detection of air showers. KASCADE and its extension
KASCADE-Grande stopped finally the active data acquisition of all its components including
the radio EAS experiment LOPES end of 2012 after more than 20 years of data taking. 
In a first release, with KCDC we
provide to the public the measured and reconstructed parameters of more than 160 million air
showers. In addition, KCDC provides the conceptional design, how the data can be treated
and processed so that they are also usable outside the community of experts in the
research field. Detailed educational examples make a use also possible for high-school
students and early stage researchers.
\end{abstract}

\section{Introduction}

\begin{figure}[h!]
\centering
\hspace*{1cm} \includegraphics[width=0.65\linewidth]{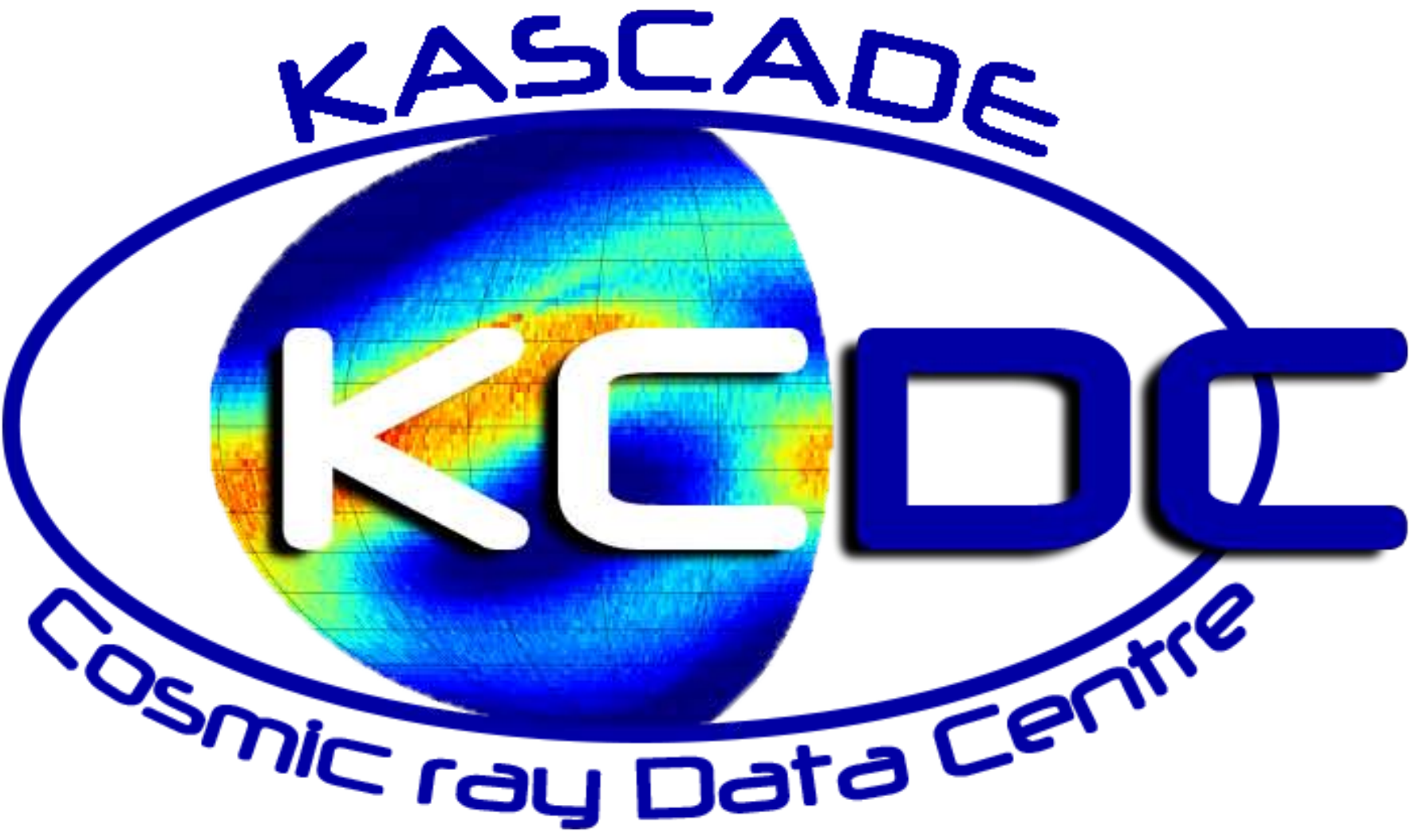}
\caption{Logo of KCDC. The link to KCDC is: \large{\bf \url{https://kcdc.ikp.kit.edu}}.}  
\label{fig_logo}
\end{figure}

\begin{figure}[h!]
\centering
\includegraphics[width=0.54\linewidth]{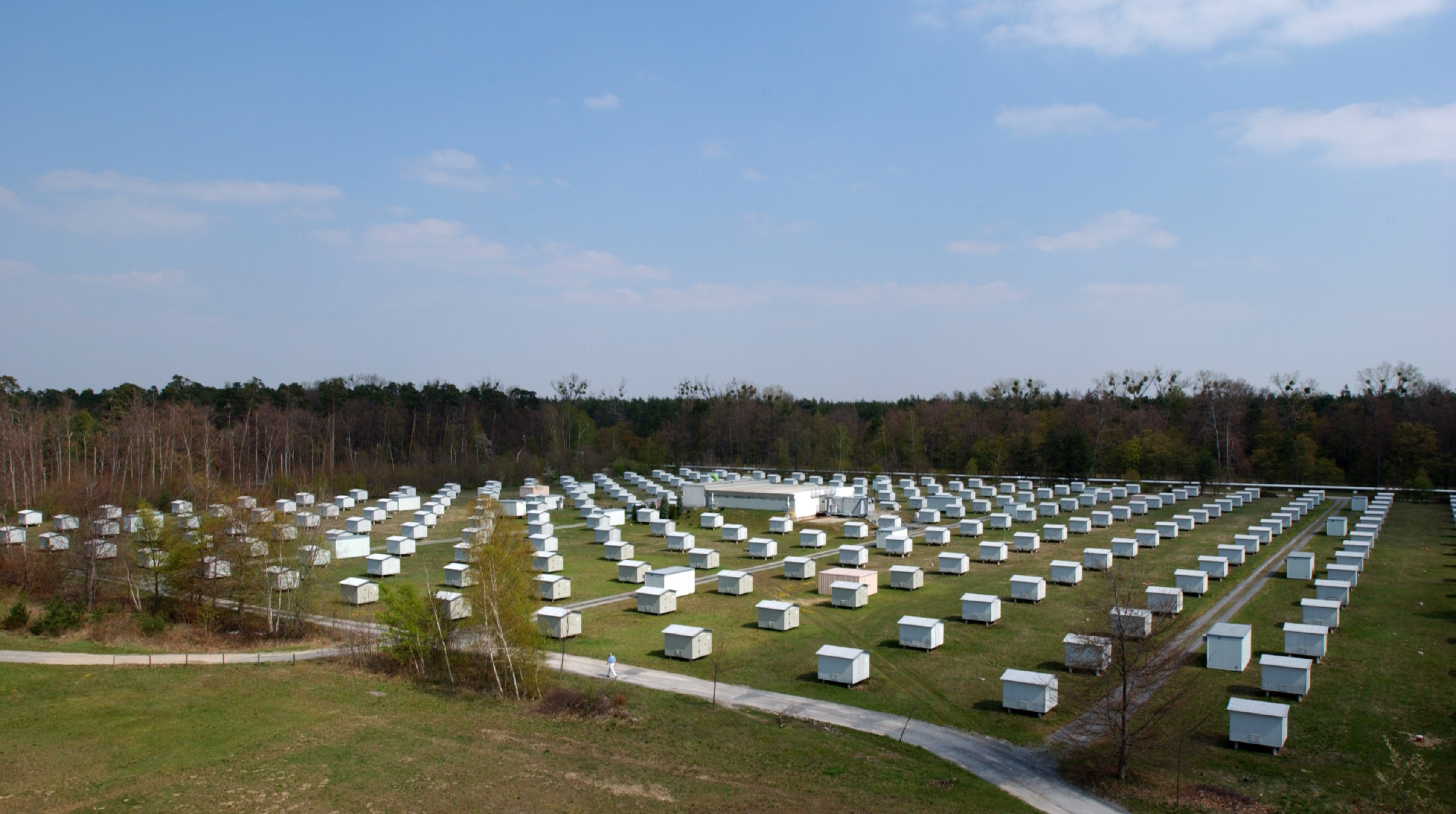}
\includegraphics[width=0.45\linewidth]{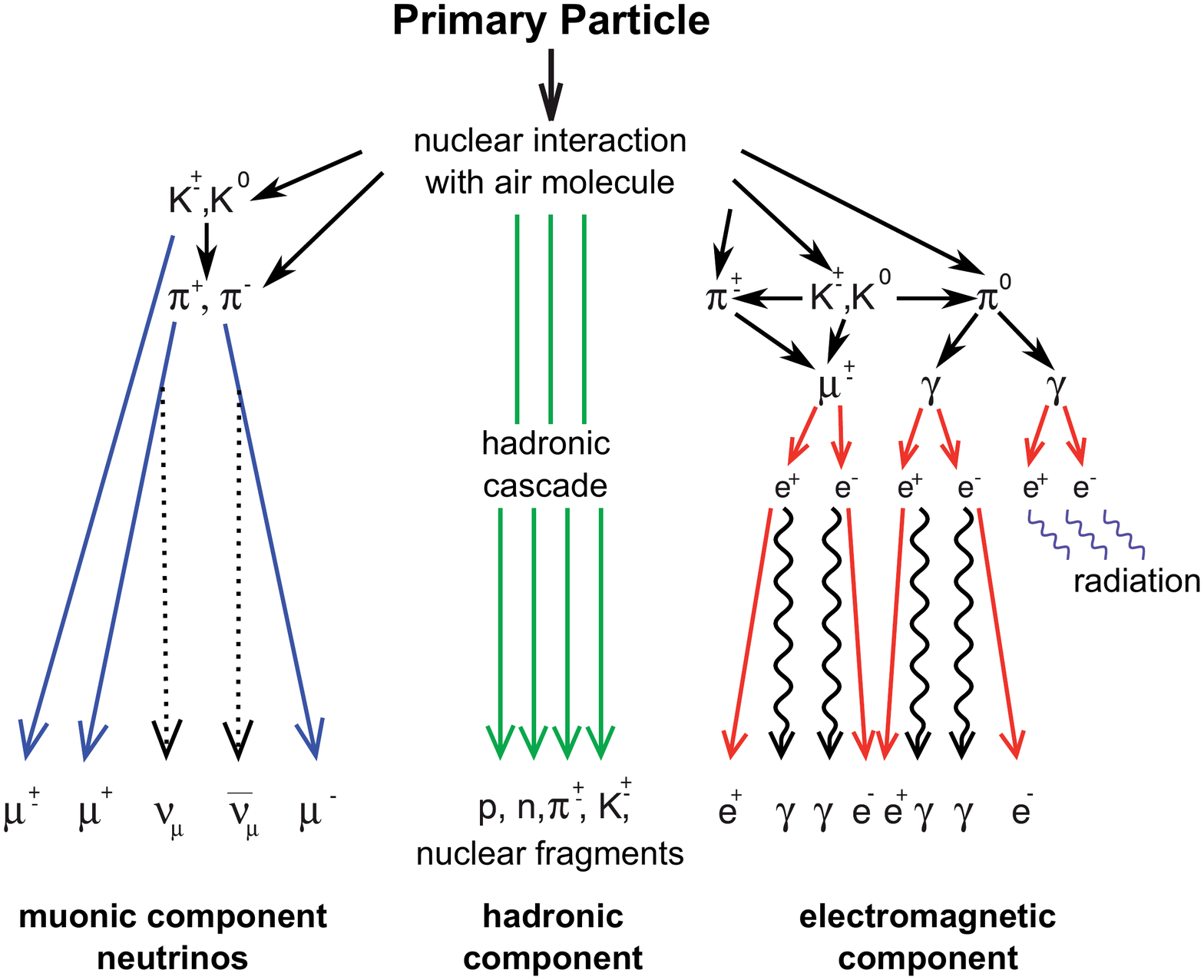}
\caption{Photograph of the KASCADE experiment (left panel); Schematic view of an extensive air shower (EAS), where KASCADE is measuring the hadronic, muonic, and electron components (right panel).}  
\label{fig_kascade}
\end{figure}

The aim of the project KCDC -- KASCADE Cosmic-ray Data Centre~\cite{kcdc} is the installation 
and establishment 
of a public data centre for high-energy astroparticle physics based on the data of the 
KASCADE experiment (logo of KCDC, see~Figure~\ref{fig_logo}). 
KASCADE (Figure~\ref{fig_kascade}) was a quite successful large detector array for measuring high-energy cosmic rays 
via the detection of extensive air showers (EAS).  KASCADE recorded data during more than 20 years on 
site of the KIT, Campus North, Karlsruhe, Germany (formerly Forschungszentrum Karlsruhe) at 
$49.1^\circ$N, $8.4^\circ$E, and $110\,$m a.s.l. 
KASCADE collected within its lifetime more than 1.7 billion events of which some 425.000.000 survived 
all quality cuts. 
Initially, about 160 million events had been made available at KCDC for public usage. 

KASCADE~\cite{kascade}, as an extensive air shower (EAS) experiment, studies the cosmic ray primary composition and 
the hadronic interactions in the energy range of $E_0 = 10^{14} - 10^{17}\,$eV. With its extension to 
KASCADE-Grande in 2003~\cite{kg-NIM10}, the range could be extended to $10^{18}\,$eV. 
EAS are generated when high-energy cosmic particles enter the Earth's atmosphere. 
Forward-boosted secondary particles as well as emitted light during the development of the EAS in 
various frequency ranges form the detectable products.

Main parts of the experiment were the Grande array spread over an area of $700 \times 700\,$m$^2$, 
the original KASCADE array covering $200 \times 200\,$m$^2$ with non-shielded and shielded 
detectors, a large-size hadron calorimeter, and additional muon tracking devices. 
The radio antenna field LOPES~\cite{lopes} and the microwave experiment CROME~\cite{crome}, 
as well as some smaller test experiments and monitoring equipment completed the experimental 
set-up of KASCADE-Grande.

One of the main results obtained by KASCADE is a picture of increasingly heavier composition above the 
`knee' in the cosmic-ray energy spectrum, caused by a break in the spectrum of the light components. 
Conventional acceleration and propagation models predict a change of the composition towards heavier 
components due to the charge dependent cut-offs in the flux of the individual elements~\cite{kas-unf,prl107}. 
The discovery of the knee in the heavy components by KASCADE-Grande, convincingly supports these 
theories~\cite{ECRS-schoo}. 

KASCADE-Grande finally stopped the active data acquisition of all its components end of 
2012 and is now decommissioned. The collaboration, however, continues the detailed 
analysis of nearly 20 years of high-quality air-shower data. 
Moreover, with KCDC, we provide to the public the edited data via a custom-made web page.

\section{KCDC in a Nutshell} 
\label{sec:kcdc}
The KASCADE/KASCADE-Grande experiment was a large-area detector for the measurement of 
cosmic ray air showers financed by taxes. 
The aim of KCDC is the installation and establishment of a public 
data centre for high-energy astroparticle physics. 
In the research field of astroparticle physics, such a data release is a novelty, whereas the data 
publication in astronomy has been established for a long time. 
Therefore, there are no completed concepts, how the data can be treated and processed so that 
they are reasonably usable outside the collaboration.
The first goal of KCDC is to make the data from the KASCADE experiment available to the community.

A concept for this kind of data centre (software and hardware) is meanwhile developed, 
implemented, and already released as a public beta version to external users. 
However, the project faces thereby still open questions, e.g. how to ensure a consistent calibration, 
how to deal with data filtering and how to provide the data in a portable format as well as how a
sustainable storage solution can be implemented. 
In addition, access rights and license policy play a non-negligible role and are considered in details.
Readers are invited to visit KCDC under 
%\begin{verbatim}https://kcdc.ikp.kit.edu/\end{verbatim} 
\verb https://kcdc.ikp.kit.edu . 

Already with the first release, KCDC provides efforts to fulfill following three basic requirements:
\begin{itemize}  
\item {\bf KCDC as data provider:}
There is free and unlimited open access to KASCADE cosmic ray data, where a selection of fully 
calibrated and reconstructed quantities per individual air shower is provided.
The access has to rely on a reliable data source with a guaranteed data quality.
\item {\bf KCDC as information platform:}
For a meaningful usage of KCDC, a detailed experiment description as well as sufficient 
meta information on the provided data is needed for any kind of data analysis.
This is accompanied by a reasonable description of the physics background as well as tutorials, 
which are focused on a level for teachers and pupils 
(in the present version of KCDC the tutorials are provided in German, only).
\item {\bf KCDC as long-term digital data archive:}
To constitute a sustainable piece of work, KCDC serves also as archive of software and data
for the collaboration as well as for the public.
\end{itemize}

\section{The Web Portal}

\begin{figure}[hb]
\centering
\includegraphics[width=0.91\linewidth]{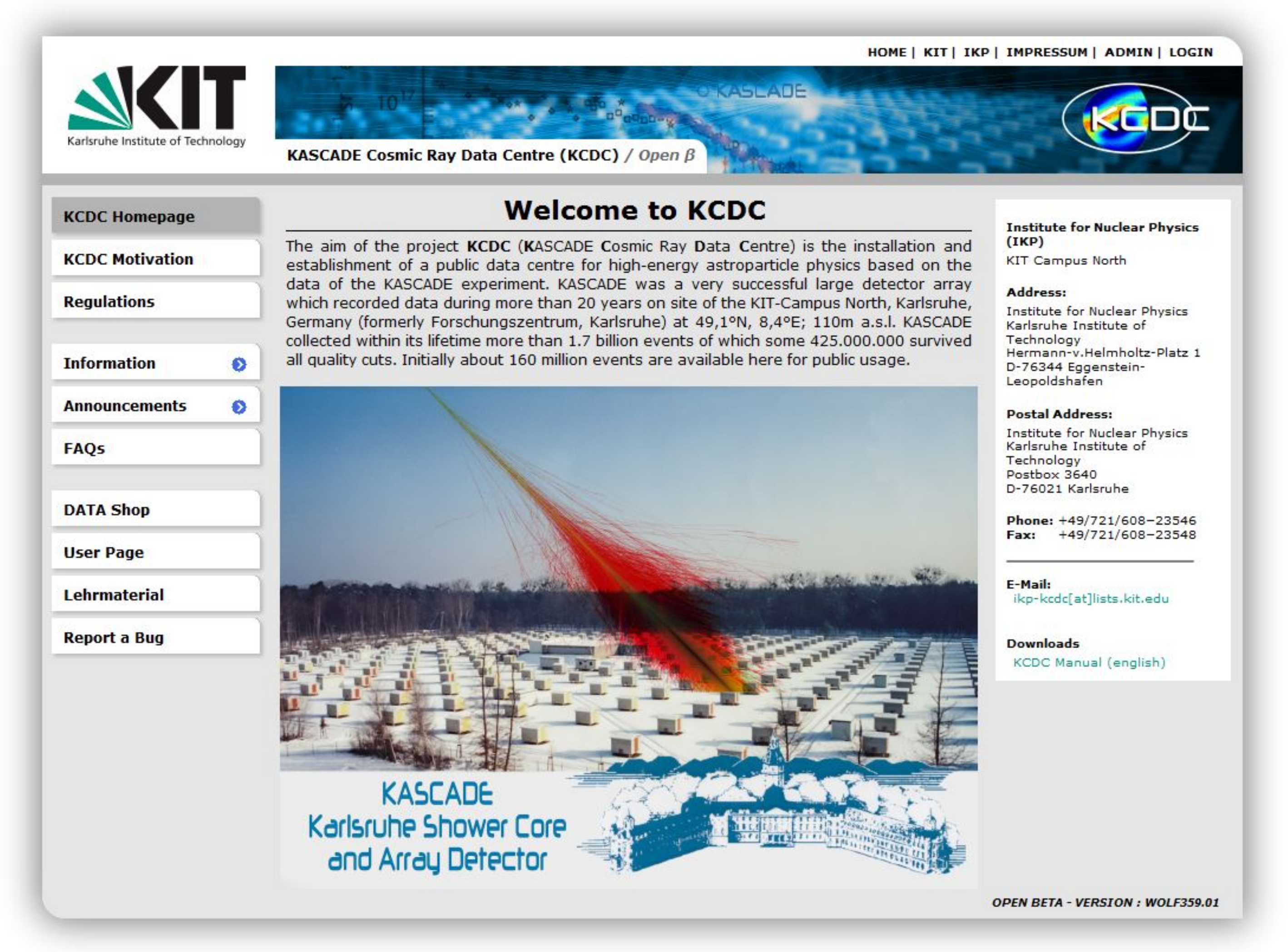}
\caption{Title page of the website of KCDC, the KASCADE Cosmic-ray Data centre.}  
\label{fig_title}
\end{figure}

The web portal (entrance page see Figure~\ref{fig_title}) as interface between the data archive, the data centre's software and the user is 
one of the most important parts of KCDC. 
It provides the door to the open data publication, where the baseline concept follows the 
`Berlin Declaration on Open Data and Open Access'~\cite{berlin} which explicitly requests the 
use of web technologies and free, unlimited access for everyone.

We declared both, the scientific and the non-scientific audience as focus of possible users. This  
requires extensive documentation of experiment, data, and software on a level understandable and 
handy for all.

\begin{figure}[ht]
\centering
\includegraphics[width=0.71\linewidth]{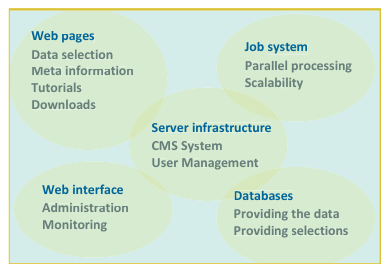}
\caption{Scheme and key words of the KCDC web portal.}  
\label{fig_portal}
\end{figure}

The portal uses modern technologies, including standard internet access and interactive data selections. 
The selected data are provided for download via a corresponding ftp-server.
Figure~\ref{fig_portal} shows schematically the basic concept of the KCDC web portal.  

It is foreseen that the software behind the data centre including the web portal is also made 
available at a later stage. Therefore, we were anxious that KCDC provides a modern software solution 
justified to perform both, publishing the KASCADE data and understandable for a general audience.
If KCDC is running successfully and is accepted by the community the software will be released as 
Open Source for free use also by other experiments. 
To serve as a general software solution for open access to (astroparticle) data, KCDC is build as a
modular, flexible framework with a good scalability (e.g.~to large computing centres).
The configuration is hold to be simple and doable also via a web interface; the entire software is 
based solely on Open Source Software (Python, Django, HTML/Javascript, CSSdata provider, etc.)

\section{Data Availability}

Since November 2013, the first release of KCDC, more than 160 million events of the
KASCADE experiment with 14 parameters per event are available.
In the first year of operation nearly 100 users registered, where we recognized access to KCDC 
by IP-addresses from more than 30 countries distributed over 5 continents.

\begin{figure}[ht]
\centering
\includegraphics[width=0.65\linewidth]{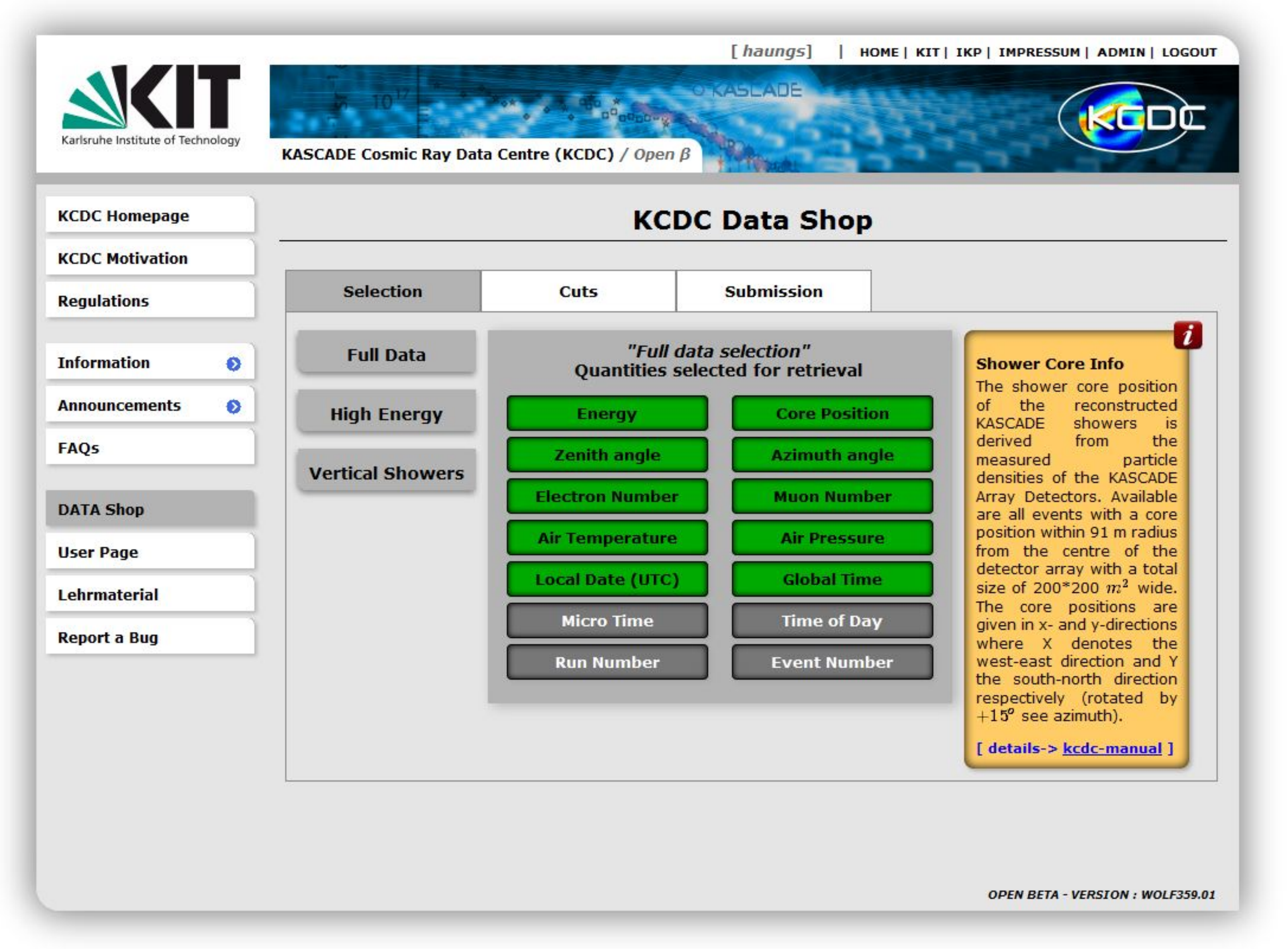}
\caption{Screen shot of the `data shop' of KCDC.}  
\label{fig_shop}
\end{figure}

Figure~\ref{fig_shop} shows a screen shot of the data shop with the list of parameters available per event. 
If registered via the `user page', the user is able to enter the data shop. 
A registration is necessary in order to ensure that the 
`End User License Agreement' is read, i.e.~the legal aspects of public data are understood (see also next chapter). 
At the data shop the user can select specific event samples. For each parameter 
a description is available in a corresponding info box appearing by a mouse-over function.
After defining cuts the selection can be submitted. The user gets an Email notification when the 
selection has been processed and is ready for download via an ftp server. 
The data will be at the user's disposal in ASCII-format including a detailed header with descriptions 
of the selection and the data format. 
Also several pre-selections of KASCADE data are available directly at the data shop.

The data can be used for any analysis, presumed that the user accept the `limited use licence' 
(following text is taken from the KCDC-EULA, see also next chapter):

{\it Subject to your agreement and continuing compliance with the KCDC Terms, 
KIT hereby grants to you a limited, personal, nonexclusive, non-transferable, 
non-assignable and fully revocable license to --
(a) use the webportal and
(b) download and use the scientific data of the KCDC in compliance with good scientific practice --
provided through the webportal or related online services for your non-commercial scientific purposes only. 
Commercial purposes are defined as projects for your own or third parties for which you are paid 
or granted values in lieu of cash for the use of the data.}

There is no restriction on the kind of analysis with the provided data nor the publication of the results.
However, the KCDC team would acknowledge notification on a use exceeding private education, as well as bug 
reports or suggestions for improvements. This can be done directly via the web portal and/or per Email to  
{ikp-kcdc@lists.kit.edu}.

\section{Legal Aspects of KCDC}

Opposite to software open source publications, there is no standard procedure yet available for 
open data publication. 
In cooperation with KIT and its law department we developed an own license based on the 
EULA (end user license agreement) model~\cite{eula}, adapted from that one often used for software. 

We had to consider a twofold issue as the license is needed for the web portal and the data. 
The KCDC approach is based on the EULA model, because it is	flexible and adaptable to our needs, it 
includes the idea of requiring a good scientific practice, and it can be signed 
during registration and can be shipped with each data package.

In our custom-made adaption of the KCDC EULA we followed some key points from industry,
like
\begin{enumerate}	
\item no warranty for damage by owner of web portal or data; 
\item no guarantee for availability or uptime of the server; 
\item	in case of disputes with local laws the EULA intention is conserved; 
\item changes are possible at any time;
\item the termination of EULA is at our digression, only,
\end{enumerate}
as well as obvious requirements from the open data idea, like
\begin{enumerate}	
\item free access to the data and the web portal;
\item good scientific practice for the work with the data; 
\item commercial usage of the data is not prohibited\footnote{Please note, that the apparent contradiction to the statement above concerning the non-commercial use of the data, is solved by following statement in the EULA: `As an alternative to this EULA, KIT offers a license to use the DATA commercially as well, on the basis of a commercial license agreement. If YOU are interested in such a license please contact KIT, Institute for Nuclear Physics (IKP), the KCDC Group and the contact person provided on the WEBPORTAL'.}; 
\item the citation of collaboration, KIT, and the web portal is mandatory;
\item free redistribution of data `as is'.
\end{enumerate}

\section{Tutorials}

The goal of having detailed tutorials, i.e.~an `education portal', is to provide the data 
also to a general public in the sense of a visible outreach of astroparticle physics.
The first tutorials are already available, but presently in German, only. English and also 
other languages can be added without any problem and will be done in the near future. 
\begin{figure}[ht]
\centering
\includegraphics[width=0.71\linewidth]{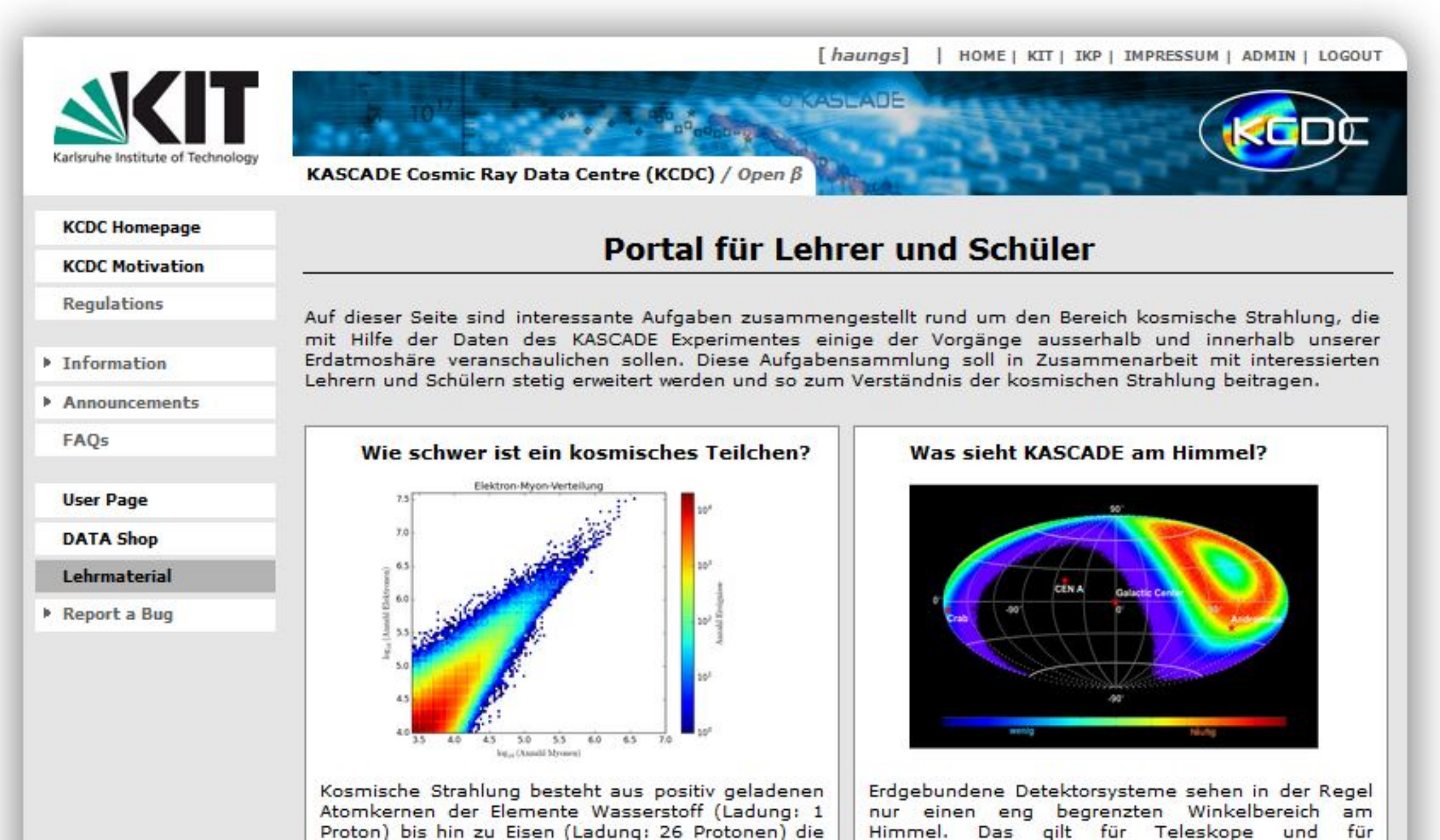}
\caption{Impression of the tutorial section of KCDC.}  
\label{fig_tuto}
\end{figure}

The target of the tutorials are teacher and pupils in high schools. This means a tutorial has 
to provide 
\begin{enumerate}
\item a basic knowledge on KASCADE, astrophysics and related topics; 
\item the required software and KCDC data (preferably as a pre-selection); 
\item a step by step explanation of a simple data analysis; 
\item a modern programming language code example; 
\item the interpretation and discussion	of the outcome.     
\end{enumerate}

This part we develop together with local teachers and pupils. 

\section{Future Steps}

Several tasks and ideas are on the working list for further development of KCDC 
(some of them are agreed on, others are under discussion):
\begin{itemize}
\item Extending the educational portal with more examples and in various languages
\item Improving the data selection process to be more comfortable and faster
\item Improving the capabilities of the data-base and corresponding software
\item Adding more data of KASCADE per individual event
\item Adding more events from KASCADE
\item Adding simulated data
\item Adding data of KASCADE-Grande and LOPES
\item Adding data of other cosmic ray experiments 
\item Publication of the software behind KCDC
\item Inclusion of KCDC in long-term data archive networks, e.g. Re3data~\cite{re3data},
\item ....
\end{itemize}

\section*{Notice}
Please note that since this article has been written, KCDC was further developed and improved. 
The description here is based on the {\it Open Beta Version: Wolf359.01}. Please visit the KCDC 
page and click the button `Developer News' for more information.

\ack
%{Acknowledgement}
The KCDC team acknowledges the continuous support of the project by the `Helmholtz Alliance 
for Astroparticle Physics HAP' funded by the Initiative and Networking Fund of the Helmholtz 
Association. In addition we acknowledge the fruitful cooperation with Hartmut Aichert, 
teacher at the local Thomas-Mann-Gymnasium, Blankenloch and his pupils.

\end{document}